\documentclass[11pt]{article}%
\usepackage{amsmath}
\usepackage{graphicx}%
\usepackage{amsfonts}%
\usepackage{amssymb}
\newcommand{\eqnb}{\begin{equation}}
\newcommand{\eqne}{\end{equation}}

\newtheorem{The}{Theorem}

\newtheorem{Pro}{Proposition}
\newtheorem{Rem}{Remark}

\begin{document}

\title{\textbf{A Matrix-Analytic Solution for Randomized Load Balancing Models with
Phase-Type Service Times}}
\author{Quan-Lin Li${}^{1}$ \hspace{0.1in}John C.S. Lui${}^{2}$ \hspace{0.1in}Wang
Yang${}^{3}$\\
${}^{1}$ School of Economics and Management Sciences \\
Yanshan University, Qinhuangdao 066004, China \\
${}^{2}$ Department of Computer Science \& Engineering \\
The Chinese University of Hong Kong, Shatin, N.T, Hong Kong\\
${}^{3}$ Institute of Network Computing \& Information Systems\\
Peking University, China }
\date{}
\maketitle

\begin{abstract}
In this paper, we provide a matrix-analytic solution for randomized load
balancing models (also known as \emph{supermarket models}) with phase-type
(PH) service times. Generalizing the service times to the phase-type
distribution makes the analysis of the supermarket models more difficult and
challenging than that of the exponential service time case which has been
extensively discussed in the literature. We first describe the supermarket
model as a system of differential vector equations, and provide a doubly
exponential solution to the fixed point of the system of differential vector
equations. Then we analyze the exponential convergence of the current
location of the supermarket model to its fixed point. Finally, we present
numerical examples to illustrate our approach and show its effectiveness in
analyzing the randomized load balancing schemes with non-exponential service
requirements.
\end{abstract}

\section{Introduction}

In the past few years, a number of companies (e.g., Amazon, Google, ,...etc)
are offering the \emph{cloud computing} service to enterprises. Furthermore,
many content publishers and application service providers are increasingly
using \emph{Data Centers} to host their services. This emerging computing
paradigm allows service providers and enterprises to concentrate on
developing and providing their own services/goods without worrying about
computing system maintenance or upgrade, and thereby significantly reduce
their operating cost. For companies that offer cloud computing service in
their data centers, they can take advantage of the variation of computing
workloads from these customers and achieve the computational multiplexing
gain. One of the important technical challenges that they have to address is
how to utilize these computing resources in the data center efficiently
since many of these servers can be virtualized. There is a growing interest
to examine simple and robust load balancing strategies to efficiently
utilize the computing resource of the server farms.

Distributed load balancing strategies, in which individual job (or customer)
decisions are based on information on a limited number of other processors,
have been studied analytically by Eager, Lazokwska and Zahorjan \cite%
{Eag:1986a,Eag:1986b,Eag:1988} and through trace-driven simulations by Zhou %
\cite{Zhou:1988}. Further, randomized load balancing is a simple and
effective mechanism to fairly utilize computing resources, and also can
deliver surprisingly good performance measures such as reducing collisions,
waiting times, backlogs,... etc. In a supermarket model, each arriving job
randomly picks a small subset of servers and examines their instantaneous
workload, and the job is routed to the least loaded server. When a job is
committed to a server, jockeying is not allowed and each server uses the
first-come-first-service (FCFS) discipline to process all jobs, e.g., see
Mitzenmacher \cite{Mit:1996a,Mit:1996b}. For the supermarket models, most of
recent research applied density dependent jump Markov processes to deal with
the simple case with Poisson arrival processes and exponential service
times, and illustrated that there exists a fixed point which decreases
doubly exponentially. Readers may refer to, such as, a simple supermarket
model by \cite{Azar:1999,Vve:1996,Mit:1996a,Mit:1996b}; simple variations by %
\cite{Mit:1998,Mit:1998a,Mit:1999a,Mit:2001,Voc:1999,Mit:2001a,Vve:1997};
load information by \cite{Mir:1989,Dah:1999,Mit:2000,Mit:2001a}; fast
Jackson network by Martin and Suhov \cite{Mar:1999,Mar:2001,Suh:2002}; and
general service times by Bramson, Lu and Prabhakar \cite{Bra:2010}. When the
arrival processes or the service times are more general, the available
results of the supermarket models are few up to now. The purpose of this
paper is to provide a novel approach for studying a supermarket model with
PH service times, and show that the fixed point decreases doubly
exponentially.

The remainder of this paper is organized as follows. In the next section, we
describe the supermarket model with the PH service times as a system of
differential vector equations based on the density dependent jump Markov
processes. In Section 3, we set up a system of nonlinear equations satisfied
by the fixed point, provide a doubly exponential solution to the system of
nonlinear equations, and compute the expected sojourn time of any arriving
customer. In Section 4, we study the exponential convergence of the current
location of the supermarket model to its fixed point. In Section 5,
numerical examples illustrate that our approach is effective in analyzing
the supermarket models from non-exponential service time requirements. Some
concluding remarks are given in Section 6.

\section{Supermarket Model}

In this section, we describe a supermarket model with the PH service times
as a system of differential vector equations based on the density dependent
jump Markov processes.

Let us formally describe the supermarket model, which is abstracted as a
multi-server multi-queue stochastic system. Customers arrive at a queueing
system of $n>1$ servers as a Poisson process with arrival rate $n\lambda$
for $\lambda>0$. The service times of these customers are of phase type with
irreducible representation $\left( \alpha,T\right) $ of order $m$. Each
arriving customer chooses $d \geq 1$ servers independently and uniformly at
random from these $n$ servers, and waits for service at the server which
currently contains the fewest number of customers. If there is a tie,
servers with the fewest number of customers will be chosen randomly. All
customers in every server will be served in the FCFS manner. Please see
Figure 1 for an illustration.

\begin{figure}[ptbh]
\centering       \includegraphics[width=5cm]{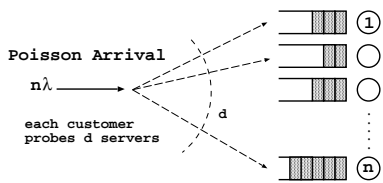}
\caption{The supermarket model: each customer can probe the loading of $d$
servers}
\label{figure: model}
\end{figure}

For the supermarket models, the PH distribution allows us to model more
realistic systems and understand their performance implication under the
randomized load balancing strategy. As indicated in \cite{harchol97}, the
process lifetime of many parallel jobs, in particular, jobs to data centers,
tends to be non-exponential. For the PH service time distribution, we use
the following irreducible representation: $\left( \alpha ,T\right) $ of
order $m$, the row vector $\alpha $ is a probability vector whose $j$th
entry is the probability that a service begins in phase $j$ for $1\leq j\leq
m$; $T$ is an $m\times m$ matrix whose $\left( i,j\right) ^{th}$ entry is
denoted by $t_{i,j}$ with $t_{i,i}<0$ for $1\leq i\leq m$, and $t_{i,j}\geq
0 $ for $1\leq i,j\leq m$ and $i\neq j$. Let $T^{0}=-Te\gvertneqq 0$, where $%
e$ is a column vector of ones with a suitable dimension in the context. The
expected service time is given by $1/\mu =-\alpha T^{-1}e$. Unless we state
otherwise, we assume that all random variables defined above are
independent, and that the system is operating in the stable region $\rho
=\lambda /\mu <1$.

We define $n_{k}^{\left( i\right) }\left( t\right) $ as the number of queues
with at least $k$ customers and the service time in phase $i$ at time $t\geq
0$. Clearly, $0\leq n_{k}^{\left( i\right) }\left( t\right) \leq n$ for $%
k\geq 0$ and $1\leq i\leq m$. Let%
\begin{equation*}
X_{n}^{\left( 0\right) }\left( t\right) =\frac{n}{n}=1,
\end{equation*}%
and $k\geq 1$
\begin{equation*}
X_{n}^{\left( k,i\right) }\left( t\right) =\frac{n_{k}^{\left( i\right)
}\left( t\right) }{n},
\end{equation*}%
which is the fraction of queues with at least $k$ customers and the service
time in phase $i$ at time $t\geq 0$. We write%
\begin{equation*}
X_{n}^{\left( k\right) }\left( t\right) =\left( X_{n}^{\left( k,1\right)
}\left( t\right) ,X_{n}^{\left( k,2\right) }\left( t\right) ,\ldots
,X_{n}^{\left( k,m\right) }\left( t\right) \right) ,\text{ \ }k\geq 1,
\end{equation*}%
\begin{equation*}
X_{n}\left( t\right) =\left( X_{n}^{\left( 0\right) }\left( t\right)
,X_{n}^{\left( 1\right) }\left( t\right) ,X_{n}^{\left( 2\right) }\left(
t\right) ,\ldots \right) .
\end{equation*}%
The state of the supermarket model may be described by the vector $%
X_{n}\left( t\right) $ for $t\geq 0$. Since the arrival process to the
queueing system is Poisson and the service times of each server are of phase
type, the stochastic process $\left\{ X_{n}\left( t\right) ,t\geq 0\right\} $
describing the state of the supermarket model is a Markov process whose
state space is given by%
\begin{eqnarray*}
\Omega _{n} &=&\{\left( g_{n}^{\left( 0\right) },g_{n}^{\left( 1\right)
},,g_{n}^{\left( 2\right) }\ldots \right) :g_{n}^{\left( 0\right)
}=1,g_{n}^{\left( k-1\right) }\geq g_{n}^{\left( k\right) }\geq 0, \\
&&\text{and \ \ }ng_{n}^{\left( k\right) }\text{ \ is a vector of
nonnegative integers for }k\geq 1\}.
\end{eqnarray*}%
Let%
\begin{equation*}
s_{0}\left( n,t\right) =E\left[ X_{n}^{\left( 0\right) }\left( t\right) %
\right]
\end{equation*}%
and $k\geq 1$%
\begin{equation*}
s_{k}^{\left( i\right) }\left( n,t\right) =E\left[ X_{n}^{\left( k,i\right)
}\left( t\right) \right] .
\end{equation*}%
Clearly, $s_{0}\left( n,t\right) =1$. We write%
\begin{equation*}
S_{k}\left( n,t\right) =\left( s_{k}^{\left( 1\right) }\left( n,t\right)
,s_{k}^{\left( 2\right) }\left( n,t\right) ,\ldots ,s_{k}^{\left( m\right)
}\left( n,t\right) \right) ,\text{ \ }k\geq 1.
\end{equation*}

As shown in Martin and Suhov \cite{Mar:1999} and Luczak and McDiarmid \cite%
{Luc:2006}, the Markov process $\left\{ X_{n}\left( t\right) ,t\geq
0\right\} $ is asymptotically deterministic as $n\rightarrow \infty $. Thus
the limits $\lim_{n\rightarrow \infty }E\left[ X_{n}^{\left( 0\right)
}\left( t\right) \right] $ and $\lim_{n\rightarrow \infty }E\left[
X_{n}^{\left( k,i\right) }\right] $ always exist by means of the law of
large numbers. Based on this, we write%
\begin{equation*}
S_{0}\left( t\right) =\lim_{n\rightarrow \infty }s_{0}\left( n,t\right) =1,
\end{equation*}%
for $k\geq 1$%
\begin{equation*}
s_{k}^{\left( i\right) }\left( t\right) =\lim_{n\rightarrow \infty
}s_{k}^{\left( i\right) }\left( n,t\right) ,
\end{equation*}%
\begin{equation*}
S_{k}\left( t\right) =\left( s_{k}^{\left( 1\right) }\left( t\right)
,s_{k}^{\left( 2\right) }\left( t\right) ,\ldots ,s_{k}^{\left( m\right)
}\left( t\right) \right)
\end{equation*}%
and%
\begin{equation*}
S\left( t\right) =\left( S_{0}\left( t\right) ,S_{1}\left( t\right)
,S_{2}\left( t\right) ,\ldots \right) .
\end{equation*}%
Let $X\left( t\right) =\lim_{n\rightarrow \infty }X_{n}\left( t\right) $.
Then it is easy to see from the Poisson arrivals and the PH service times
that $\left\{ X\left( t\right) ,t\geq 0\right\} $ is also a Markov process
whose state space is given by%
\begin{equation*}
\Omega =\left\{ \left( g^{\left( 0\right) },g^{\left( 1\right) },g^{\left(
2\right) },\ldots \right) :g^{\left( 0\right) }=1,g^{\left( k-1\right) }\geq
g^{\left( k\right) }\geq 0\right\} .
\end{equation*}%
If the initial distribution of the Markov process $\left\{ X_{n}\left(
t\right) ,t\geq 0\right\} $ approaches the Dirac delta-measure concentrated
at a point $g\in $ $\Omega $, then its steady-state distribution is
concentrated in the limit on the trajectory $S_{g}=\left\{ S\left( t\right)
:t\geq 0\right\} $. This indicates a law of large numbers for the time
evolution of the fraction of queues of different lengths. Furthermore, the
Markov process $\left\{ X_{n}\left( t\right) ,t\geq 0\right\} $ converges
weakly to the fraction vector $S\left( t\right) =\left( S_{0}\left( t\right)
,S_{1}\left( t\right) ,S_{2}\left( t\right) ,\ldots \right) $, or for a
sufficiently small $\varepsilon >0$,%
\begin{equation*}
\lim_{n\rightarrow \infty }P\left\{ ||X_{n}\left( t\right) -S\left( t\right)
||\geq \varepsilon \right\} =0,
\end{equation*}%
where $||a||$ is the $L_{\infty }$-norm of vector $a$.

In what follows we provide a system of differential vector equations in
order to determine fraction vector $S\left( t\right) $. To that end, we
introduce the \emph{Hadamard Product} of two matrices $A=\left(
a_{i,j}\right) $ and $B=\left( b_{i,j}\right) $ as follows:
\begin{equation*}
A\odot B=\left( a_{i,j}b_{i,j}\right) .
\end{equation*}%
Specifically, for $k\geq 2$, we have%
\begin{equation*}
A^{\odot k}=\underset{k\text{ matrix }A}{\underbrace{A\odot A\odot \cdots
\odot A}}.
\end{equation*}

To determine the fraction vector $S\left( t\right) $, we need to set up a
system of differential vector equations satisfied by $S\left( t\right) $ by
means of the density dependent jump Markov process. To that end, we provide
a concrete example for $k\geq 2$ to indicate how to derive the the system of
differential vector equations.

Consider the supermarket model with $n$ servers, and determine the expected
change in the number of queues with at least $k$ customers over a small time
period of length d$t$. The probability vector that during this time period,
any arriving customer joins a queue of size $k-1$ is given by%
\begin{equation*}
n\left[ \lambda S_{k-1}^{\odot d}\left( n,t\right) -\lambda S_{k}^{\odot
d}\left( n,t\right) \right] \text{d}t.
\end{equation*}%
Similarly, the probability vector that a customer leaves a server queued by $%
k$ customers is given by%
\begin{equation*}
n\left[ S_{k}\left( n,t\right) T+S_{k+1}\left( n,t\right) T^{0}\alpha \right]
\text{d}t.
\end{equation*}%
Therefore we can obtain%
\begin{align*}
\text{d}E\left[ n_{k}\left( n,t\right) \right] =& n\left[ \lambda
S_{k-1}^{\odot d}\left( n,t\right) -\lambda S_{k}^{\odot d}\left( n,t\right) %
\right] \text{d}t \\
& +n\left[ S_{k}\left( n,t\right) T+S_{k+1}\left( n,t\right) T^{0}\alpha %
\right] \text{d}t,
\end{align*}%
which leads to%
\begin{equation*}
\frac{\text{d}S_{k}\left( n,t\right) }{\text{d}t}=\lambda S_{k-1}^{\odot
d}\left( n,t\right) -\lambda S_{k}^{\odot d}\left( n,t\right) +S_{k}\left(
n,t\right) T+S_{k+1}\left( n,t\right) T^{0}\alpha .
\end{equation*}%
Taking $n\rightarrow \infty $ in the both sides of Equation (\ref{Equ1}), we
have%
\begin{equation*}
\frac{\text{d}S_{k}\left( t\right) }{\text{d}t}=\lambda S_{k-1}^{\odot
d}\left( t\right) -\lambda S_{k}^{\odot d}\left( t\right) +S_{k}\left(
t\right) T+S_{k+1}\left( t\right) T^{0}\alpha .
\end{equation*}

Using a similar analysis to Equation (\ref{Equ2}), we can obtain a system of
differential vector equations for the fraction vector $S\left( t\right)
=\left( S_{0}\left( t\right) ,S_{1}\left( t\right) ,S_{2}\left( t\right)
,\ldots \right) $ as follows:%
\begin{equation*}
S_{0}\left( t\right) =1,
\end{equation*}%
\begin{equation}
\frac{\mathtt{d}}{\text{d}t}S_{0}\left( t\right) =-\lambda S_{0}^{d}\left(
t\right) +S_{1}\left( t\right) T^{0},  \label{Eq1}
\end{equation}%
\begin{equation}
\frac{\mathtt{d}}{\text{d}t}S_{1}\left( t\right) =\lambda \alpha
S_{0}^{d}\left( t\right) -\lambda S_{1}^{\odot d}\left( t\right)
+S_{1}\left( t\right) T+S_{2}\left( t\right) T^{0}\alpha ,  \label{Eq2}
\end{equation}%
and for $k\geq 2$,%
\begin{equation}
\frac{\mathtt{d}}{\text{d}t}S_{k}\left( t\right) =\lambda S_{k-1}^{\odot
d}\left( t\right) -\lambda S_{k}^{\odot d}\left( t\right) +S_{k}\left(
t\right) T+S_{k+1}\left( t\right) T^{0}\alpha .  \label{Eq3}
\end{equation}

\begin{Rem}
Mitzenmacher \cite{Mit:1996a,Mit:1996b} provided an heuristical and
interesting method to establish such systems of differential equations, but
they lack a rigorous mathematical meaning for understanding the stochastic
process $\left\{ X_{n}\left( t\right) ,t\geq 0\right\} $ in which $%
X_{n}\left( t\right) =( X_{n}^{\left( 0\right) }\left( t\right)
,X_{n}^{\left( 1\right) }\left( t\right),$ $X_{n}^{\left(
2\right) }\left( t\right),\ldots ) $ and $X_{n}^{\left( k\right)
}\left( t\right) =n_{k}\left( t\right) /n$ for $k\geq 0$. This
section, following Martin and Suhov \cite{Mar:1999} and Luczak and
McDiarmid \cite{Luc:2006}, gives some necessary mathematical
analysis for the stochastic process $\left\{ X_{n}\left( t\right) ,
t\geq 0\right\} $
and the system of differential vector equations
(\ref{Eq1}), (\ref{Eq2}) and (\ref{Eq3}).
\end{Rem}

\section{A Matrix-Analytic Solution}

In this section, we provide a doubly exponential solution to the fixed point
of the system of differential vector equations (\ref{Eq1}), (\ref{Eq2}) and (%
\ref{Eq3}).

A row vector $\pi =\left( \pi _{0},\pi _{1},\pi _{2},\ldots \right) $ is
called a fixed point of the fraction vector $S\left( t\right) $ if $%
\lim_{t\rightarrow +\infty }S\left( t\right) =\pi $. In this case, it is
easy to see that%
\begin{equation*}
\lim_{t\rightarrow +\infty }\left[ \frac{\mathtt{d}}{\text{d}t}S\left(
t\right) \right] =0.
\end{equation*}%
Therefore, as $t\rightarrow +\infty $ the system of differential vector
equations (\ref{Eq1}), (\ref{Eq2}) and (\ref{Eq3}) can be simplified as%
\begin{equation}
-\lambda \pi _{0}^{d}+\pi _{1}T^{0}=0,  \label{Eq4}
\end{equation}%
\begin{equation}
\lambda \alpha \pi _{0}^{d}-\lambda \pi _{1}^{\odot d}+\pi _{1}T+\pi
_{2}T^{0}\alpha =0,  \label{Eq5}
\end{equation}%
and for $k\geq 2$,%
\begin{equation}
\lambda \pi _{k-1}^{\odot d}-\lambda \pi _{k}^{\odot d}+\pi _{k}T+\pi
_{k+1}T^{0}\alpha =0.  \label{Eq6}
\end{equation}

In general, it is more difficult and challenging to express the fixed point
of the supermarket models with more general arrival processes or service
times, because the systems of nonlinear equations are more complicated for
computation. Fortunately, we can derive a closed-form expression for the
fixed point $\pi =(\pi _{0},\pi _{1},\pi _{2},...)$ for the supermarket
model with PH service times by means of a novel matrix-analytic approach
given as follows.

Noting that $S_{0}\left( t\right) =1$ for all $t\geq 0$, it is easy to see
that $\pi _{0}=1$. It follows from Equation (\ref{Eq4}) that%
\begin{equation}
\pi _{1}T^{0}=\lambda .  \label{Eq7}
\end{equation}%
To solve Equation (\ref{Eq7}), we denote by $\omega $ the stationary
probability vector of the irreducible Markov chain $T+T^{0}\alpha $.
Obviously, we have%
\begin{equation*}
\omega T^{0}=\mu ,
\end{equation*}%
\begin{equation}
\frac{\lambda }{\mu }\omega T^{0}=\lambda .  \label{Eq8}
\end{equation}%
Thus, we obtain $\pi _{1}=\frac{\lambda }{\mu }\omega =\rho \cdot \omega $.
Based on the fact that $\pi _{0}=1$ and $\pi _{1}=\rho \cdot \omega $, it
follows from Equation (\ref{Eq5}) that%
\begin{equation*}
\lambda \alpha -\lambda \rho ^{d}\cdot \omega ^{\odot d}+\rho \cdot \omega
T+\pi _{2}T^{0}\alpha =0,
\end{equation*}%
which leads to%
\begin{equation*}
\lambda -\lambda \rho ^{d}\cdot \omega ^{\odot d}e+\rho \cdot \omega Te+\pi
_{2}T^{0}=0.
\end{equation*}%
Note that $\omega Te=-\mu $, we obtain%
\begin{equation*}
\pi _{2}T^{0}=\lambda \rho ^{d}\omega ^{\odot d}e.
\end{equation*}%
Let $\theta =\omega ^{\odot d}e$. Then it is easy to see that $\theta \in
\left( 0,1\right) $, and%
\begin{equation*}
\pi _{2}T^{0}=\lambda \theta \rho ^{d}.
\end{equation*}%
Using a similar analysis to Equation (\ref{Eq8}), we have%
\begin{equation}
\pi _{2}=\frac{\lambda \theta \rho ^{d}}{\mu }\omega =\theta \rho
^{d+1}\cdot \omega .  \label{equa1}
\end{equation}%
Based on $\pi _{1}=\rho \cdot \omega $ and $\pi _{2}=\theta \rho ^{d+1}\cdot
\omega $, it follows from Equation (\ref{Eq6}) that for $k=2$,
\begin{equation*}
\lambda \rho ^{d}\cdot \omega ^{\odot d}-\lambda \theta ^{d}\rho
^{d^{2}+d}\cdot \omega ^{\odot d}+\theta \rho ^{d+1}\cdot \omega T+\pi
_{3}T^{0}\alpha =0,
\end{equation*}%
which leads to%
\begin{equation*}
\lambda \theta \rho ^{d}-\lambda \theta ^{d+1}\rho ^{d^{2}+d}+\theta \rho
^{d+1}\cdot \omega Te+\pi _{3}T^{0}=0,
\end{equation*}%
thus we obtain%
\begin{equation*}
\pi _{3}T^{0}=\lambda \theta ^{d+1}\rho ^{d^{2}+d}.
\end{equation*}%
Using a similar analysis on Equation (\ref{Eq8}), we have%
\begin{equation}
\pi _{3}=\frac{\lambda \theta ^{d+1}\rho ^{d^{2}+d}}{\mu }\omega =\theta
^{d+1}\rho ^{d^{2}+d+1}\cdot \omega .  \label{equa2}
\end{equation}%
Based on Equations (\ref{equa1}) and (\ref{equa2}), we may infer that there
is a structured expression $\pi _{k}=\theta ^{d^{k-2}+d^{k-3}+\cdots
+d+1}\rho ^{d^{k-1}+d^{k-2}+\cdots +d+1}\cdot \omega $, for $k\geq 1$. To
that end, the following theorem states this important result.

\begin{The}
The fixed point $\pi=\left(  \pi_{0},\pi_{1},\pi_{2},\ldots\right)  $ is
unique and is given by%
\[
\pi_{0}=1, \hspace{0.2in} \pi_{1}=\rho\cdot\omega
\]
and for $k\geq2,$%
\begin{equation}
\pi_{k}=\theta^{d^{k-2}+d^{k-3}+\cdots+1}\rho^{d^{k-1}+d^{k-2}+\cdots+1}%
\cdot\omega, \label{Eq9}%
\end{equation}
or%
\begin{eqnarray}
\pi_{k}  &  = &\theta^{\frac{d^{k-1}-1}{d-1}}\rho^{\frac{d^{k}-1}{d-1}}%
\cdot\omega\nonumber
  =  \rho^{d^{k-1}}\left(  \theta\rho\right)  ^{\frac{d^{k-1}-1}{d-1}}%
\cdot\omega.
\label{Eq10}%
\end{eqnarray}
\end{The}

\textbf{Proof:} By induction, one can easily derive the above result.

It is clear that Equation (\ref{Eq9}) is correct for the cases with $l=2,3$
according to Equations (\ref{equa1}) and (\ref{equa2}). Now, we assume that
Equation (\ref{Eq9}) is correct for the cases with $l=k$. Then it follows
from Equation (\ref{Eq6}) that for $l=k+1$, we have
\begin{align*}
\lambda & \theta ^{d^{k-2}+d^{k-3}+\cdots +d}\rho ^{d^{k-1}+d^{k-2}+\cdots
+d}\cdot \omega ^{\odot d}-\lambda \theta ^{d^{k-1}+d^{k-2}+\cdots +d}\rho
^{d^{k}+d^{k-1}+\cdots +d}\cdot \omega ^{\odot d} \\
& +\theta ^{d^{k-2}+d^{k-3}+\cdots +1}\rho ^{d^{k-1}+d^{k-2}+\cdots +1}\cdot
\omega T+\pi _{k+1}T^{0}\alpha =0,
\end{align*}%
which leads to%
\begin{align*}
\lambda & \theta ^{d^{k-2}+d^{k-3}+\cdots +d+1}\rho ^{d^{k-1}+d^{k-2}+\cdots
+d}-\lambda \theta ^{d^{k-1}+d^{k-2}+\cdots +d+1}\rho ^{d^{k}+d^{k-1}+\cdots
+d} \\
& +\theta ^{d^{k-2}+d^{k-3}+\cdots +1}\rho ^{d^{k-1}+d^{k-2}+\cdots +1}\cdot
\omega Te+\pi _{k+1}T^{0}=0,
\end{align*}%
thus we obtain%
\begin{equation*}
\pi _{k+1}T^{0}=\lambda \theta ^{d^{k-1}+d^{k-2}+\cdots +d+1}\rho
^{d^{k}+d^{k-1}+\cdots +d}.
\end{equation*}%
By a similar analysis to (\ref{Eq8}), we have%
\begin{align*}
\pi _{k+1}& =\frac{\lambda \theta ^{d^{k-1}+d^{k-2}+\cdots +d+1}\rho
^{d^{k}+d^{k-1}+\cdots +d}}{\mu }\omega \\
& =\theta ^{d^{k-1}+d^{k-2}+\cdots +d+1}\rho ^{d^{k}+d^{k-1}+\cdots
+d+1}\cdot \omega .
\end{align*}%
This completes the proof. \hspace*{\fill} \rule{1.8mm}{2.5mm}

Now, we compute the expected sojourn time $T_{d}$\ that a tagged arriving
customer spends in the supermarket model. For the PH service times, a tagged
arriving customer is the $k$th customer in the corresponding queue with
probability vector $\pi _{k-1}^{\odot d}-\pi _{k}^{\odot d}$. When $k\geq 1$%
, the head customer in the queue has been served, and so its service time is
residual and is denoted as $X_{R}$. Let $X$ be of phase type with
irreducible representation $\left( \alpha ,T\right) $. Then $X_{R}$ is of
phase type with irreducible representation $\left( \omega ,T\right) $.
Clearly, we have%
\begin{equation*}
E\left[ X\right] =\alpha \left( -T\right) ^{-1}e,\text{ \ }E\left[ X_{R}%
\right] =\omega \left( -T\right) ^{-1}e.
\end{equation*}%
Thus it is easy to see that the expected sojourn time of the tagged arriving
customer is given by%
\begin{align*}
E\left[ T_{d}\right] & =\left( \pi _{0}^{d}-\pi _{1}^{\odot d}e\right) E%
\left[ X\right] +\sum_{k=1}^{\infty }\left( \pi _{k}^{\odot d}-\pi
_{k+1}^{\odot d}\right) e\left\{ E\left[ X_{R}\right] +kE\left[ X\right]
\right\}  \\
& =\pi _{1}^{\odot d}e\left\{ E\left[ X_{R}\right] -E\left[ X\right]
\right\} +E\left[ X\right] \left[ 1+\sum_{k=1}^{\infty }\pi _{k}^{\odot d}e%
\right]  \\
& =\rho ^{d}\theta \left( \omega -\alpha \right) \left( -T\right)
^{-1}e+\alpha \left( -T\right) ^{-1}e\left( 1+\sum_{k=1}^{\infty }\theta ^{%
\frac{d^{k}-1}{d-1}}\rho ^{\frac{d^{k+1}-d}{d-1}}\right) .
\end{align*}%
When the arrival process and the service time distribution are Poisson and
exponential, respectively, it is clear that $\alpha =\omega =\theta =1$ and $%
\alpha \left( -T\right) ^{-1}e=1/\mu $, thus we have%
\begin{equation*}
E\left[ T_{d}\right] =\frac{1}{\mu }\sum_{k=0}^{\infty }\rho ^{\frac{%
d^{k+1}-d}{d-1}},
\end{equation*}%
which is the same as Corollary 3.8 in Mitzenmacher \cite{Mit:1996b}.

In what follows we consider an interesting problem: how many moments of the
service time distribution are needed to obtain a better accuracy for
computing the fixed point or the expected sojourn time. It is well-known
from the theory of probability distributions that the first three moments is
basic for analyzing such an accuracy, and we can construct a PH distribution
of order 2 by using the first three moments. Telek and Heindl \cite{Tel:2002}
provided a fitting procedure for matching a PH distribution of order 2 from
the first three moments exactly. It is necessary to list the fitting
procedure as follows:

For a nonnegative random variable $X$, let $m_{n}=E\left[ X^{n}\right] $, $%
n\geq 1$\textbf{. }We take a PH distribution of order 2 with the canonical
representation $\left( \alpha ,T\right) $, where $\mathbf{\alpha =}\left(
\eta ,1-\eta \right) $ and%
\begin{equation*}
T\mathbf{=}\left(
\begin{array}{cc}
-\xi _{1} & \xi _{1} \\
0 & -\xi _{2}%
\end{array}%
\right) ,
\end{equation*}%
$0\leq \eta \leq 1$ and $0<\xi _{1}\leq \xi _{2}$. Note that the three
unknown parameters $\eta $, $\xi _{1}$ and $\xi _{2}$ can be obtained from
the first three moments $m_{1}$, $m_{2}$ and $m_{3}$ of an arbitrary general
distribution.

\begin{table}[tbp]
\caption{Specific Bounds of the First Three Moments}\renewcommand{%
\arraystretch}{1.3} \centering
\begin{tabular}{|c|c|c|}
\hline
\bfseries Moment & \bfseries Condition & \bfseries Bounds \\ \hline
$m_{1}$ &  & $0<m_{1}<\infty $ \\ \hline
$m_{2}$ &  & $1.5m_{1}^{2}\leq m_{2}$ \\ \hline
$m_{3}$ & $0.5\leq c_{X}^{2}\leq 1$ & $3m_{1}^{3}\left( 3c_{X}^{2}-1+\sqrt{2}%
\left( 1-c_{X}^{2}\right) ^{\frac{3}{2}}\right) \leq m_{3}\leq
6m_{1}^{3}c_{X}^{2}$ \\ \hline
& $1<c_{X}^{2}$ & $\frac{3}{2}m_{1}^{3}\left( 1+c_{X}^{2}\right)
^{2}<m_{3}<\infty $ \\ \hline
\end{tabular}%
\end{table}

In Table 1, $c_{X}^{2}=m_{2} \diagup m_{1}^{2}-1$ is the squared coefficient
of variation. If the moments do not satisfy these conditions in Table 1,
then we may analyze the following four cases:

(a.1) if $m_{2}<1.5m_{1}^{2}$, then we take $m_{2}=1.5m_{1}^{2}$;

(a.2) if $0.5\leq c_{X}^{2}\leq 1$, and $m_{3}<3m_{1}^{3}\left( 3c_{X}^{2}-1
+\sqrt{2}\left( 1-c_{X}^{2}\right) ^{\frac{3}{2}}\right)$, then we take $%
m_{3}=3m_{1}^{3}\left( 3c_{X}^{2}-1+\sqrt{2}\left( 1-c_{X}^{2}\right) ^{%
\frac{3}{2}}\right)$;

(a.3) if $0.5\leq c_{X}^{2}\leq 1$, and $m_{3}>6m_{1}^{3}c_{X}^{2}$, then we
take $m_{3}=6m_{1}^{3}c_{X}^{2}$; and

(a.4) if $1<c_{X}^{2}$, and $m_{3}\leq \frac{3}{2} m_{1}^{3}\left(
1+c_{X}^{2}\right) ^{2}$, then we take $m_{3}=\frac{3}{2}m_{1}^{3}\left(
1+c_{X}^{2}\right) ^{2}$.

Let $c=3m_{2}^{2}-2m_{1}m_{3}$, $d=2m_{1}^{2}-m_{2}$, $b=3m_{1}m_{2}-m_{3}$
and $a=b^{2}-6cd$. If the moments respectively satisfy their specific bounds%
\emph{\ }shown in Table 1 or the exceptive four cases, then three unknown
parameters $\eta $, $\xi _{1}$ and $\xi _{2}$ can be computed in the
following three cases.

(1) If $c>0$, then%
\begin{equation*}
\eta =\frac{-b+6m_{1}d+\sqrt{a}}{b+\sqrt{a}},\text{ }\xi _{1}=\frac{b-\sqrt{a%
}}{c},\text{ }\xi _{2}=\frac{b+\sqrt{a}}{c}.
\end{equation*}

(2) If $c<0$, then%
\begin{equation*}
\eta =\frac{b-6m_{1}d+\sqrt{a}}{-b+\sqrt{a}},\text{ }\xi _{1}=\frac{b+\sqrt{a%
}}{c},\text{ }\xi _{2}=\frac{b-\sqrt{a}}{c}.
\end{equation*}

(3) If $c=0$, then%
\begin{equation*}
\eta =0,\text{ }\xi _{1}>0,\text{ }\xi _{2}=\frac{1}{m_{1}}.
\end{equation*}

From the above discussion, we can always construct a PH distribution of
order 2 to approximate an arbitrary general distribution with the same first
three moments. In fact, such an approximation achieves a better accuracy in
computation.

For the PH distribution of order 2, we have%
\begin{equation*}
T+T^{0}\alpha =\left(
\begin{array}{cc}
-\xi _{1} & \xi _{1} \\
0 & -\xi _{2}%
\end{array}%
\right) +\left(
\begin{array}{c}
0 \\
\xi _{2}%
\end{array}%
\right) \left(
\begin{array}{cc}
\eta  & 1-\eta
\end{array}%
\right) =\left(
\begin{array}{cc}
-\xi _{1} & \xi _{1} \\
\xi _{2}\eta  & -\xi _{2}\eta
\end{array}%
\right) ,
\end{equation*}%
which leads to%
\begin{equation*}
\omega =\left( \frac{\xi _{2}\eta }{\xi _{1}+\xi _{2}\eta },\frac{\xi _{1}}{%
\xi _{1}+\xi _{2}\eta }\right)
\end{equation*}%
and%
\begin{equation*}
\theta =\frac{\xi _{1}^{d}+\xi _{2}^{d}\eta ^{d}}{\left( \xi _{1}+\xi
_{2}\eta \right) ^{d}}.
\end{equation*}%
Thus, the PH distribution of order 2 can effectively approximates an
arbitrary general service time distribution in the supermarket model under
the same first three moments, and specifically, all the computations are
very simple to implement.

\begin{Rem}
Bramson, Lu and Prabhakar \cite{Bra:2010} provided a modularized
program based on ansatz for treating the supermarket model with a
general service time distribution. They organized a functional
equation $\pi =F\left( G\left( \pi \right) \right)$ for analyzing
the stationary probability vector $\pi$ in terms of insensitivity
and generalized Fibonacci sequences, although the operators $F$ and
$G$ are not easy to be given for this supermarket model. This paper studies the supermarket model with a PH service time
distribution, provides the doubly exponential solution to the fixed
point, and is specifically related to the phase type environment by
means of the crucial factor $\theta =\omega ^{\odot d}e$. Note that
the PH distributions are dense in the set of all nonnegative random
variables, this paper can numerically provide necessary
understanding for the role played by the general service time
distribution in performance analysis of the supermarket model by means of the PH approximation of order 2.
\end{Rem}

\section{Exponential convergence to the fixed point}

In this section, we study the exponential convergence of the current
location $S\left( t\right)$ of the supermarket model to its fixed point $\pi$%
.

For the supermarket model, the initial point $S\left( 0\right) $ can affect
the current location $S\left( t\right) $ for each $t>0$, since the service
process in the supermarket model is under a unified structure. To that end,
we provide some notation for comparison of two vectors. Let $a=\left(
a_{1},a_{2},a_{3},\ldots \right) $ and $b=\left( b_{1},b_{2},b_{3},\ldots
\right) $. We write $a\prec b$ if $a_{k}<b_{k}$ for some $k\geq 1$ and $%
a_{l}\leq b_{l}$ for $l\neq k,l\geq 1$; and $a\preceq b$ if $a_{k}\leq b_{k}$
for all $k\geq 1$. Now, we can obtain the following useful proposition whose
proof is clear from a sample path analysis and thus is omitted here.

\begin{Pro}
\label{Prop1}If $S\left(  0\right)  \preceq\widetilde{S}\left(  0\right)  $,
then $S\left(  t\right)  \preceq\widetilde{S}\left(  t\right)  $.
\end{Pro}

Based on Proposition \ref{Prop1}, the following theorem shows that the fixed
point $\pi $ is an upper bound of the current location $S\left( t\right) $
for all $t\geq 0$.

\begin{The}
For the supermarket model, if there exists some $k$ such that $S_{k}\left(
0\right)  =0$, then the sequence $\left\{  S_{k}\left(  t\right)  \right\}  $
has an upper bound sequence which decreases doubly exponentially for all
$t\geq0 $, that is, $S\left(  t\right)  \preceq\pi$ for all $t\geq0$.
\end{The}

\textbf{Proof:} \ Let $\widetilde{S}_{k}\left( 0\right) =\pi _{k}$ for $%
k\geq 1$. Then for each $k\geq 1$, $\widetilde{S}_{k}\left( t\right) =%
\widetilde{S}_{k}\left( 0\right) =\pi _{k}$ for all $t\geq 0$, since $%
\widetilde{S}\left( 0\right) =\left( \widetilde{S}_{1}\left( 0\right) ,%
\widetilde{S}_{2}\left( 0\right) ,\widetilde{S}_{2}\left( 0\right) ,\ldots
\right) $ is a fixed point in the supermarket model. If $S_{k}\left(
0\right) =0$ for some $k$, then $S_{k}\left( 0\right) \prec \widetilde{S}%
_{k}\left( 0\right) $ and $S_{j}\left( 0\right) \preceq \widetilde{S}%
_{j}\left( 0\right) $\ for $j\neq k,j\geq 1$, thus $S\left( 0\right) \preceq
\widetilde{S}\left( 0\right) $. It is easy to see from Proposition \ref%
{Prop1} that $S_{k}\left( t\right) \preceq \widetilde{S}_{k}\left( t\right)
=\pi _{k}$ for all $k\geq 1$ and $t\geq 0$. Thus we obtain that for all $%
k\geq 1$ and $t\geq 0$
\begin{equation*}
S_{k}\left( t\right) \leq \theta ^{\frac{d^{k-1}-1}{d-1}}\rho ^{\frac{d^{k}-1%
}{d-1}}\cdot \omega .
\end{equation*}%
This completes the proof. \hspace*{\fill} \rule{1.8mm}{2.5mm}

To show the exponential convergence, we define a Lyapunov function $\Phi
\left( t\right) $ as
\begin{equation*}
\Phi \left( t\right) =\sum_{k=1}^{\infty }w_{k}\left[ \pi _{k}-S_{k}\left(
t\right) \right] e
\end{equation*}%
in terms of the fact that $S_{k}\left( t\right) \preceq \pi _{k}$ for $k\geq
1$ and $\pi _{0}=S_{0}\left( t\right) =1$, where $\left\{ w_{k}\right\} $ is
a positive scalar sequence with $w_{k+1}\geq w_{k}\geq w_{1}=1$ for $k\geq 2$%
.

The following theorem measures the distance $\Phi \left( t\right) $ of the
current location $S\left( t\right) $ for $t\geq 0$ to the fixed point $\pi $%
, and illustrates that this distance between the fixed point and the current
location is very close to zero with exponential convergence. This shows that
from a suitable starting point, the supermarket model can be quickly close
to the fixed point.

\begin{The}
For $t\geq0$, $\Phi\left(  t\right)  \leq c_{0}e^{-\delta t}$, where
$c_{0}$ and $\delta$ are two positive constants. In this case, the
potential function $\Phi\left(  t\right)  $ is exponentially
convergent.
\end{The}

\textbf{Proof:} \ Note that
\begin{equation*}
\Phi \left( t\right) =\sum_{k=1}^{\infty }w_{k}\left[ \pi _{k}-S_{k}\left(
t\right) \right] e,
\end{equation*}%
we have
\begin{equation*}
\frac{d}{dt}\Phi \left( t\right) =-\sum_{k=1}^{\infty }w_{k}\frac{d}{dt}%
S_{k}\left( t\right) e.
\end{equation*}%
It follows from Equations (\ref{Eq1}) to (\ref{Eq3}) that%
\begin{align*}
\frac{d}{dt}\Phi \left( t\right) =& -w_{1}[\lambda S_{0}^{d}\left( t\right)
\alpha -\lambda S_{1}^{\odot d}\left( t\right) +S_{1}\left( t\right)
T+S_{2}\left( t\right) T^{0}\alpha ]e \\
& -\sum_{k=1}^{\infty }w_{k}[\lambda S_{k-1}^{\odot d}\left( t\right)
-\lambda S_{k}^{\odot d}\left( t\right) +S_{k}\left( t\right)
T+S_{k+1}\left( t\right) T^{0}\alpha ]e.
\end{align*}%
By means of $S_{0}\left( t\right) =1$ and $Te=-T^{0}$, we can obtain%
\begin{align}
\frac{d}{dt}\Phi \left( t\right) =& -w_{1}[\lambda -\lambda S_{1}^{\odot
d}\left( t\right) e-S_{1}\left( t\right) T^{0}+S_{2}\left( t\right) T^{0}]
\notag \\
& -\sum_{k=2}^{\infty }w_{k}[\lambda S_{k-1}^{\odot d}\left( t\right)
e-\lambda S_{k}^{\odot d}\left( t\right) e-S_{k}\left( t\right)
T^{0}+S_{k+1}\left( t\right) T^{0}].  \label{Equ3}
\end{align}%
We take some nonnegative constants $c_{k}\left( t\right) $ and $d_{k}\left(
t\right) $ for $k\geq 1$ such that%
\begin{equation*}
\lambda =f_{1}\left( t\right) S_{1}\left( t\right) T^{0},
\end{equation*}%
for $k\geq 1$%
\begin{equation*}
\lambda S_{k}^{\odot d}\left( t\right) e=c_{k}\left( t\right) \left[ \pi
_{k}-S_{k}\left( t\right) \right] e
\end{equation*}%
and%
\begin{equation*}
S_{k}\left( t\right) T^{0}=d_{k}\left( t\right) \left[ \pi _{k}-S_{k}\left(
t\right) \right] e.
\end{equation*}%
Then it follows from (\ref{Equ3}) that%
\begin{align*}
\frac{d}{dt}\Phi \left( t\right) & =-\left\{ \left[ \left(
w_{2}-w_{1}\right) \right] c_{1}\left( t\right) +w_{1}\left[ f_{1}\left(
t\right) -1\right] d_{1}\left( t\right) \right\} \cdot \left[ \pi
_{1}-S_{1}\left( t\right) \right] e \\
& -\sum_{k=2}^{\infty }\left[ \left( w_{k+1}-w_{k}\right) c_{k}\left(
t\right) +\left( w_{k-1}-w_{k}\right) d_{k}\left( t\right) \right] \cdot %
\left[ \pi _{k}-S_{k}\left( t\right) \right] e.
\end{align*}%
For a constant $\delta >0$, we take%
\begin{equation*}
w_{1}=1,
\end{equation*}%
\begin{equation*}
\left[ \left( w_{2}-w_{1}\right) \right] c_{1}\left( t\right) +w_{1}\left[
f_{1}\left( t\right) -1\right] d_{1}\left( t\right) \geq \delta w_{1}
\end{equation*}%
and%
\begin{equation*}
\left( w_{k+1}-w_{k}\right) c_{k}\left( t\right) +\left(
w_{k-1}-w_{k}\right) d_{k}\left( t\right) \geq \delta w_{k}.
\end{equation*}%
In this case, it is easy to see that%
\begin{equation*}
w_{2}\geq 1+\frac{\delta +1-f_{1}\left( t\right) }{c_{1}\left( t\right) }
\end{equation*}%
and for $k\geq 2$%
\begin{equation*}
w_{k+1}\geq w_{k}+\frac{\delta w_{k}}{c_{k}\left( t\right) }+\frac{%
d_{k}\left( t\right) }{c_{k}\left( t\right) }\left( w_{k}-w_{k-1}\right) .
\end{equation*}%
Thus we have
\begin{equation*}
\frac{d}{dt}\Phi \left( t\right) \leq -\delta \sum_{k=0}^{\infty }w_{k}\left[
\pi _{k}-S_{k}\left( t\right) \right] e=-\delta \Phi \left( t\right) ,
\end{equation*}%
which can leads to%
\begin{equation*}
\Phi \left( t\right) \leq c_{0}e^{-\delta t}.
\end{equation*}%
This completes the proof. \hspace*{\fill} \rule{1.8mm}{2.5mm}

\section{Numerical examples}

In this section, we provide some numerical examples to illustrate that our
approach is effective and efficient in the study of supermarket models with
non-exponential service requirements, including Erlang service time
distributions, hyper-exponential service time distributions and PH service
time distributions.

\noindent \textbf{Example one} (Erlang Distribution) We consider an $m$%
-order Erlang distribution with the irreducible PH representation $%
(\alpha,T) $, where$\alpha=\left( 1,0,\ldots,0,0\right)$ and
\begin{equation*}
T=\left(
\begin{array}{ccccc}
-\eta & \eta &  &  &  \\
& -\eta & \eta &  &  \\
&  & \ddots & \ddots &  \\
&  &  & -\eta & \eta \\
&  &  &  & -\eta%
\end{array}
\right) ,\text{ \ \ }T^{0}=\left(
\begin{array}{c}
0 \\
0 \\
\vdots \\
0 \\
\eta%
\end{array}
\right) .
\end{equation*}
It is clear that%
\begin{equation*}
T+T^{0}\alpha=\left(
\begin{array}{ccccc}
-\eta & \eta &  &  &  \\
& -\eta & \eta &  &  \\
&  & \ddots & \ddots &  \\
&  &  & -\eta & \eta \\
\eta &  &  &  & -\eta%
\end{array}
\right) ,
\end{equation*}
which leads to the stationary probability vector of the Markov chain $%
T+T^{0}\alpha$ as follows:
\begin{equation*}
\omega=\left( \frac{1}{m},\frac{1}{m},\ldots\frac{1}{m},\frac{1}{m}\right);
\hspace{0.1in} \mu=\omega T^{0}=\frac{\eta}{m}; \hspace{0.1in} \rho=\frac{%
\lambda}{\mu}=\frac{m\lambda}{\eta}; \hspace{0.1in} \theta=m\left( \frac{1}{m%
}\right) ^{d}=m^{1-d}.
\end{equation*}%
Thus we obtain%
\begin{align*}
\pi_{k} & =m^{1-d^{k-1}}\left( \frac{m\lambda}{\eta}\right) ^{\frac{d^{k}-1}{%
d-1}}\left( \frac{1}{m},\frac{1}{m},\ldots\frac{1}{m},\frac{1}{m}\right) \\
& =m^{\frac{d^{k-1}+d-2}{d-1}}\left( \frac{\lambda}{\eta}\right) ^{\frac{%
d^{k}-1}{d-1}}\left( \frac{1}{m},\frac{1}{m},\ldots\frac{1}{m},\frac{1}{m}%
\right) .
\end{align*}

Let $\lambda =1$. If $\rho =\frac{m\lambda }{\eta }<1$, then this
supermarket model is stable. In the stable case, $\eta >m$. We may consider
the following simple cases: \newline
\noindent (a) If $m=2$ and $d=2$, then $\pi _{k}=2^{2^{k-1}}\eta ^{1-2^{k}}$.

\noindent (b) If $m=3$ and $d=2$, then $\pi_{k}=3^{2^{k-1}}\eta^{1-2^{k}}$.

Based on the two simple examples with $\lambda =1$ and $d=2$, we need to
illustrate how the fixed point depends on the stage number $m$ and the
exponential service rate $\eta $. To that end, we write $\pi _{k}\left(
m,\eta \right) $. It is easy to see that for a given pair $\left( k,\eta
\right) $ for $\eta >m$ and $k=1,2,\ldots ,$ we have%
\begin{equation*}
\pi _{k}\left( 1,\eta \right) <\pi _{k}\left( 2,\eta \right) <\cdots <\pi
_{k}\left( m,\eta \right) <\cdots .
\end{equation*}%
On the other hand, for a given pair $\left( k,m\right) $ for $m,k=1,2,\ldots
,$ we can see that $\pi _{k}\left( m,\eta \right) $ is a decreasing function
of $\eta $.

Let us consider the average response time of the supermarket model with an $%
m-$stage Erlang distribution. We first consider a parallel system with $%
n=100 $ servers and the service time distribution is exponential. We
normalize the average service time to unity and vary the arrival rate $%
\lambda $. For the $m-$stage Erlang distribution, the bigger the number $m$
is, the bigger its variance is. Table \ref{table: simulation result 1}
illustrates the average response time under different probe size $d$. One
can observe that there is a dramatic improvement (or reduction) in the
average response time when increasing the probe size $d$.

\begin{table}[htb]
\caption{Average response time for exponential service time}
\label{table: simulation result 1}\centering {\scriptsize
\begin{tabular}{|c||c|c|c|}
\hline
\textbf{number of servers} ($n$) & \textbf{probe size} ($d$) & \textbf{%
arrival rate ($\lambda$)} & \textbf{response time} ($E[\mathcal{T}]$) \\
\hline\hline
100 & 2 & 0.500000 & 1.395977 \\ \hline
100 & 2 & 0.700000 & 1.768194 \\ \hline
100 & 2 & 0.800000 & 2.072020 \\ \hline
100 & 2 & 0.900000 & 2.721852 \\ \hline\hline
100 & 3 & 0.500000 & 1.395320 \\ \hline
100 & 3 & 0.700000 & 1.604113 \\ \hline
100 & 3 & 0.800000 & 1.802933 \\ \hline
100 & 3 & 0.900000 & 2.209601 \\ \hline\hline
100 & 5 & 0.900000 & 1.916280 \\ \hline
\end{tabular}
}
\end{table}

We further analyze the cases that the service time is either distributed
according to $2$-stage Erlang or $3$-stage Erlang distribution. Similarly,
we normalized the total average service time as unity and we vary the
arrival rate $\lambda $. Tables \ref{table: simulation result 2} and \ref%
{table: simulation result 3} illustrate the average response time under
different probe size $d$. One can observe that

\begin{itemize}
\item Simple probing size $d$ can significantly improve the performance by
lowering the average response time.

\item When the service time has lower variance, the average response time is
lower.
\end{itemize}

\begin{table}[htb]
\caption{Average response time for $2-$stage Erlang service time}
\label{table: simulation result 2}\centering {\scriptsize
\begin{tabular}{|c||c|c|c|}
\hline
\textbf{number of servers} ($n$) & \textbf{probe size} ($d$) & \textbf{%
arrival rate ($\lambda$)} & \textbf{response time} ($E[\mathcal{T}]$) \\
\hline\hline
100 & 2 & 0.500000 & 1.353783 \\ \hline
100 & 2 & 0.700000 & 1.599851 \\ \hline
100 & 2 & 0.800000 & 1.829199 \\ \hline
100 & 2 & 0.900000 & 2.298470 \\ \hline\hline
100 & 3 & 0.500000 & 1.325610 \\ \hline
100 & 3 & 0.700000 & 1.492651 \\ \hline
100 & 3 & 0.800000 & 1.639987 \\ \hline
100 & 3 & 0.900000 & 1.941196 \\ \hline\hline
100 & 5 & 0.900000 & 1.739867 \\ \hline
\end{tabular}
}
\end{table}

\begin{table}[htb]
\caption{Average response time for $3-$stage Erlang service time}
\label{table: simulation result 3}\centering {\scriptsize
\begin{tabular}{|c||c|c|c|}
\hline
\textbf{number of servers} ($n$) & \textbf{probe size} ($d$) & \textbf{%
arrival rate ($\lambda$)} & \textbf{response time} ($E[\mathcal{T}]$) \\
\hline\hline
100 & 2 & 0.500000 & 1.322544 \\ \hline
100 & 2 & 0.700000 & 1.539621 \\ \hline
100 & 2 & 0.800000 & 1.739972 \\ \hline
100 & 2 & 0.900000 & 2.148191 \\ \hline\hline
100 & 3 & 0.500000 & 1.298863 \\ \hline
100 & 3 & 0.700000 & 1.452785 \\ \hline
100 & 3 & 0.800000 & 1.581663 \\ \hline
100 & 3 & 0.900000 & 1.834704 \\ \hline\hline
100 & 5 & 0.900000 & 1.678233 \\ \hline
\end{tabular}
}
\end{table}

\noindent \textbf{Example two} (Hyper-Exponential Distribution) We consider
an $m$-order hyper-exponential distribution $F\left( x\right)
=1-\sum\limits_{k=1}^{m}\alpha _{k}\exp \left\{ -\eta _{k}x\right\} $, or
the probability density function $f\left( x\right)
=\sum\limits_{k=1}^{m}\alpha _{k}\eta _{k}\exp \left\{ -\eta _{k}x\right\} $%
. It is clear that the hyper-exponential distribution is of phase type with
the irreducible representation $(\alpha ,T)$, where $\alpha =\left( \alpha
_{1},\alpha _{2},\ldots ,\alpha _{m}\right) $, and
\begin{equation*}
T=\left(
\begin{array}{cccc}
-\eta _{1} &  &  &  \\
& -\eta _{2} &  &  \\
&  & \ddots &  \\
&  &  & -\eta _{m}%
\end{array}%
\right) ,\text{ \ }T^{0}=\left(
\begin{array}{c}
\eta _{1} \\
\eta _{2} \\
\vdots \\
\eta _{m}%
\end{array}%
\right) ,
\end{equation*}%
which lead to%
\begin{equation*}
T+T^{0}\alpha =\left(
\begin{array}{cccc}
-\eta _{1}\left( 1-\alpha _{1}\right) & \eta _{1}\alpha _{2} & \cdots & \eta
_{1}\alpha _{m} \\
\eta _{2}\alpha _{1} & -\eta _{2}\left( 1-\alpha _{2}\right) & \cdots & \eta
_{2}\alpha _{m} \\
\vdots & \vdots &  & \vdots \\
\eta _{m}\alpha _{1} & \eta _{m}\alpha _{2} & \cdots & -\eta _{m}\left(
1-\alpha _{m}\right)%
\end{array}%
\right) .
\end{equation*}%
In general, the system of equations $\omega \left( T+T^{0}\alpha \right) =0$
and $\omega e=1$ does not admit a simple analytic solution. For a convenient
description, we only consider a simple one with $m=2$. In this case, we
obtain%
\begin{equation*}
\omega =\left( \frac{\alpha _{1}\eta _{2}}{\alpha _{1}\eta _{2}+\alpha
_{2}\eta _{1}},\frac{\alpha _{2}\eta _{1}}{\alpha _{1}\eta _{2}+\alpha
_{2}\eta _{1}}\right) ,\hspace{0.2in}\mu =\frac{\eta _{1}\eta _{2}\left(
\alpha _{1}+\alpha _{2}\right) }{\alpha _{1}\eta _{2}+\alpha _{2}\eta _{1}},
\end{equation*}%
\begin{equation*}
\rho =\frac{\lambda }{\mu }=\frac{\lambda \left( \alpha _{1}\eta _{2}+\alpha
_{2}\eta _{1}\right) }{\eta _{1}\eta _{2}\left( \alpha _{1}+\alpha
_{2}\right) },\hspace{0.1in}\theta =\left( \frac{\alpha _{1}\eta _{2}}{%
\alpha _{1}\eta _{2}+\alpha _{2}\eta _{1}}\right) ^{d}+\left( \frac{\alpha
_{2}\eta _{1}}{\alpha _{1}\eta _{2}+\alpha _{2}\eta _{1}}\right) ^{d}
\end{equation*}%
and%
\begin{align*}
\pi _{k} = & \left[ \left( \frac{\alpha _{1}\eta _{2}}{\alpha _{1}\eta
_{2}+\alpha _{2}\eta _{1}}\right) ^{d}+\left( \frac{\alpha _{2}\eta _{1}}{%
\alpha _{1}\eta _{2}+\alpha _{2}\eta _{1}}\right) ^{d}\right] ^{\frac{%
d^{k-1}-1}{d-1}} \\
& \cdot \left[ \frac{\lambda \left( \alpha _{1}\eta _{2}+\alpha _{2}\eta
_{1}\right) }{\eta _{1}\eta _{2}\left( \alpha _{1}+\alpha _{2}\right) }%
\right] ^{\frac{d^{k}-1}{d-1}}\left( \frac{\alpha _{1}\eta _{2}}{\alpha
_{1}\eta _{2}+\alpha _{2}\eta _{1}},\frac{\alpha _{2}\eta _{1}}{\alpha
_{1}\eta _{2}+\alpha _{2}\eta _{1}}\right) .
\end{align*}%
Tables \ref{table: 1} and \ref{table: 2} indicate how the doubly exponential
solution ($\pi _{1}$ to $\pi _{5}$) depends on the vectors $\eta =\left(
\eta _{1},\eta _{2}\right) $ and $\alpha =\left( \alpha _{1},\alpha
_{2}\right) $, respectively.

\begin{table}[htb]
\caption{The doubly exponential solution depends on $\protect\eta$}
\label{table: 1}\centering {\scriptsize
\begin{tabular}{|c||c|c|c|}
\hline
& $\eta=(3,3)$ & $\eta=(3,10)$ & $\eta=(3,20)$ \\ \hline\hline
$\pi_{1}$ & (0.1667,\;0.1667) & (0.1667,\;0.0500) & (0.1667,\;0.0250) \\
\hline
$\pi_{2}$ & (0.0093,\;0.0093) & (0.0050,\;0.0015) & (0.0047,\;0.0007) \\
\hline
$\pi_{3}$ & (2.858e-05,\;2.858e-05) & (4.626e-06,\;1.388e-06) &
(3.819e-06,\;5.728e-07) \\ \hline
$\pi_{4}$ & (2.722e-10,\;2.722e-10) & (3.888e-12,\;1.166e-12) &
(2.485e-12,\;3.728e-13) \\ \hline
$\pi_{5}$ & (2.470e-20,\;2.470e-20) & (2.746e-24,\;8.238e-25) &
(1.053e-24,\;1.579e-25) \\ \hline
\end{tabular}
}
\end{table}

\begin{table}[htb]
\caption{The doubly exponential solution depends on $\protect\alpha$}
\label{table: 2}\centering {\scriptsize
\begin{tabular}{|c||c|c|c|}
\hline\hline
& $\alpha=(0.5,\;0.5)$ & $\alpha=(0.2,\;0.8)$ & $\alpha=(0.8,\;0.2)$ \\
\hline\hline
$\pi_{1}$ & (0.1667,\;0.1667) & (0.0667,\;0.0267) & (0.2667,\;0.0067) \\
\hline
$\pi_{2}$ & (0.0047,\;0.0005) & (0.0003,\;0.0001) & (0.0190,\;0.0005) \\
\hline
$\pi_{3}$ & (3.680e-06,\;3.680e-07) & (9.136e-09,\;3.654e-09) &
(9.607e-05,\;2.402e-06) \\ \hline
$\pi_{4}$ & (2.280e-12,\;2.280e-13) & (6.454e-18,\;2.582e-18) &
(2.463e-09,\;6.157e-11) \\ \hline
$\pi_{5}$ & (8.752e-25,\;8.752e-26) & (3.221e-36,\;1.289e-36) &
(1.618e-18,\;4.046e-20) \\ \hline
\end{tabular}
}
\end{table}

Let us consider the average response time of the supermarket model with an $%
m $-stage hyper-exponential service time distribution. We consider a
parallel system with $n=100$ servers and the probability density function of
the service time of a customer is given by%
\begin{equation*}
f(x)=0.5\times (2\times e^{-2x})+0.25\times (0.5\times e^{-0.5x})+0.25\times
(e^{-x}).
\end{equation*}%
Note that the total average service time is normalized to unity and we vary
the arrival rate $\lambda $. Table \ref{table: simulation result 4}
illustrates the average response time under different probe size $d$. One
can observe that there is a dramatic reduction in the average response time
when increasing the probe size. Furthermore, when the service time has a
higher variance (we here compare it with the exponential distribution or $m-$%
stage Erlang distribution), the average service time is much higher. This
indicates that we improve the performance of the supermarket model, one has
to increase the probe size $d$.

\begin{table}[htb]
\caption{Average response time for $3-$stage Hyper-exponential service time}
\label{table: simulation result 4}\centering {\scriptsize
\begin{tabular}{|c||c|c|c|}
\hline
\textbf{number of servers} ($n$) & \textbf{probe size} ($d$) & \textbf{%
arrival rate ($\lambda$)} & \textbf{response time} ($E[\mathcal{T}]$) \\
\hline\hline
100 & 2 & 0.500000 & 1.552282 \\ \hline
100 & 2 & 0.700000 & 1.969132 \\ \hline
100 & 2 & 0.800000 & 2.360255 \\ \hline
100 & 2 & 0.900000 & 3.225117 \\ \hline\hline
100 & 3 & 0.500000 & 1.462128 \\ \hline
100 & 3 & 0.700000 & 1.723764 \\ \hline
100 & 3 & 0.800000 & 1.947548 \\ \hline
100 & 3 & 0.900000 & 2.476718 \\ \hline\hline
100 & 5 & 0.900000 & 2.066462 \\ \hline
\end{tabular}
}
\end{table}

\noindent \textbf{Example three} (PH Distribution) We consider an $m$-order
PH distribution with irreducible representation $\left( \alpha,T\right) $.
For $m=2,d=2,\alpha=\left( 1/2,1/2\right) $ and%
\begin{equation*}
T\left( 1\right) =\left(
\begin{array}{cc}
-4 & 3 \\
2 & -7%
\end{array}
\right) ,T\left( 2\right) =\left(
\begin{array}{cc}
-5 & 3 \\
2 & -7%
\end{array}
\right) ,T\left( 3\right) =\left(
\begin{array}{cc}
-4 & 4 \\
2 & -7%
\end{array}
\right) ,
\end{equation*}
Table \ref{table: 4} illustrates how the doubly exponential solution depends
on the PH matrices $T\left( 1\right) $, $T\left( 2\right) $ and $T\left(
3\right) $, respectively.

\begin{table}[tbh]
\caption{The doubly exponential solution depends on the PH matrices $T(i)$}
\label{table: 4}\centering {\scriptsize
\begin{tabular}{|c|c|c|c|}
\hline
& $T(1)$ & $T(2)$ & $T(3)$ \\ \hline\hline
$\pi_{1}$ & (0.2045,\;0.1591) & (0.1410,\;0.1026) & (0.3125,\; 0.2500) \\
\hline
$\pi_{2}$ & (0.0137,\;0.0107) & (0.0043,\;0.0031) & (0.0500,\;0.0400) \\
\hline
$\pi_{3}$ & (6.193e-05,\;4.817e-05) & (3.965e-06,\;2.884e-06) & (0.0013
,\;0.0010) \\ \hline
$\pi_{4}$ & (1.259e-09,\;9.793e-10) & (3.390e-12,\;2.465e-12) &
(8.446e-07,\;6.757e-07) \\ \hline
$\pi_{5}$ & (5.204e-19,\;4.048e-19) & (2.478e-24,\;1.802e-24) &
(3.656e-13,\; 2.925e-13) \\ \hline
\end{tabular}
}
\end{table}

To discuss how different caused by a non-exponential distribution versus an
exponentially distributed service time with the same mean, for the above
three PH distributions we take three corresponding exponential distributions
with service rates $\mu(1) = 2.7500, \mu(2)=3.4118$ and $\mu(3)=2.3529$,
respectively. Table \ref{table: 4+} illustrates how the doubly exponential
solution ($\pi_{1}$ to $\pi_{5}$) depends on the three service rates. Since
the exponential distribution has a lower variance than the PH distribution,
it is seen from Tables \ref{table: 4} and \ref{table: 4+} that the service
time has lower variance, $\pi_k$(Exp)$<\pi_k$(PH)$e$.

\begin{table}[tbh]
\caption{The doubly exponential solution depends on exponential service
rates $\protect\mu(i)$}
\label{table: 4+}\centering {\scriptsize
\begin{tabular}{|c|c|c|c|}
\hline
& $\mu(1) = 2.7500$ & $\mu(2)=3.4118$ & $\mu(3)=2.3529$ \\ \hline\hline
$\pi_{1}$ & 0.3636 & 0.2931 & 0.4250 \\ \hline
$\pi_{2}$ & 0.0481 & 0.0252 & 0.0768 \\ \hline
$\pi_{3}$ & 8.408e-04 & 1.858e-04 & 0.0025 \\ \hline
$\pi_{4}$ & 2.571e-07 & 1.012e-08 & 2.667e-06 \\ \hline
$\pi_{5}$ & 2.402e-14 & 3.004e-17 & 3.030e-12 \\ \hline
\end{tabular}
}
\end{table}

For the PH and exponential service times, the following two figures provides
a comparison for the expected sojourn time. Clearly, the PH service time
makes the lower expected sojourn time.
\begin{figure}[tbp]
\begin{minipage}[htb]{0.5\textwidth}
\centering
\includegraphics[width=6cm]{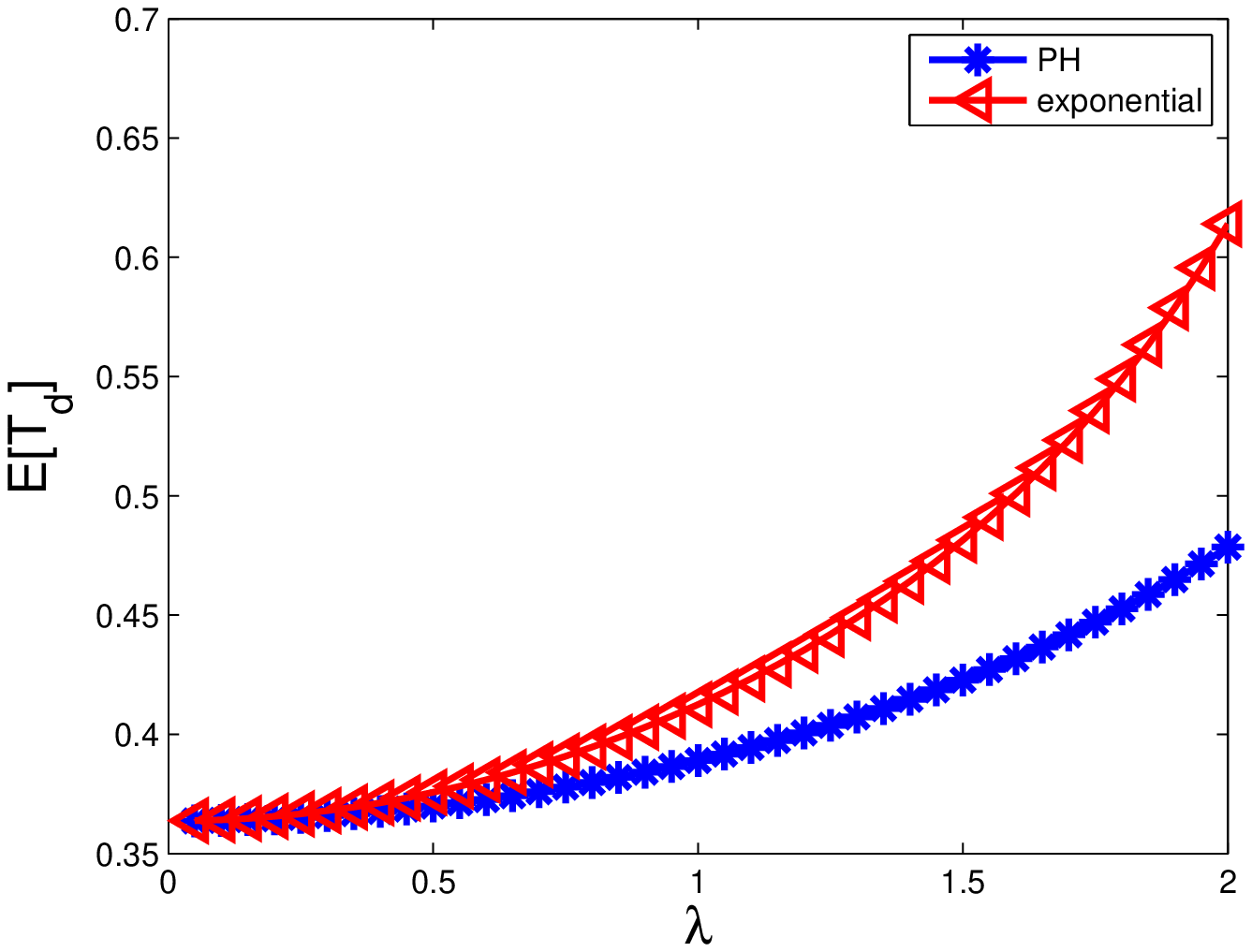}
\end{minipage}%
\begin{minipage}[htb]{0.5\textwidth}
\centering
\includegraphics[width=6cm]{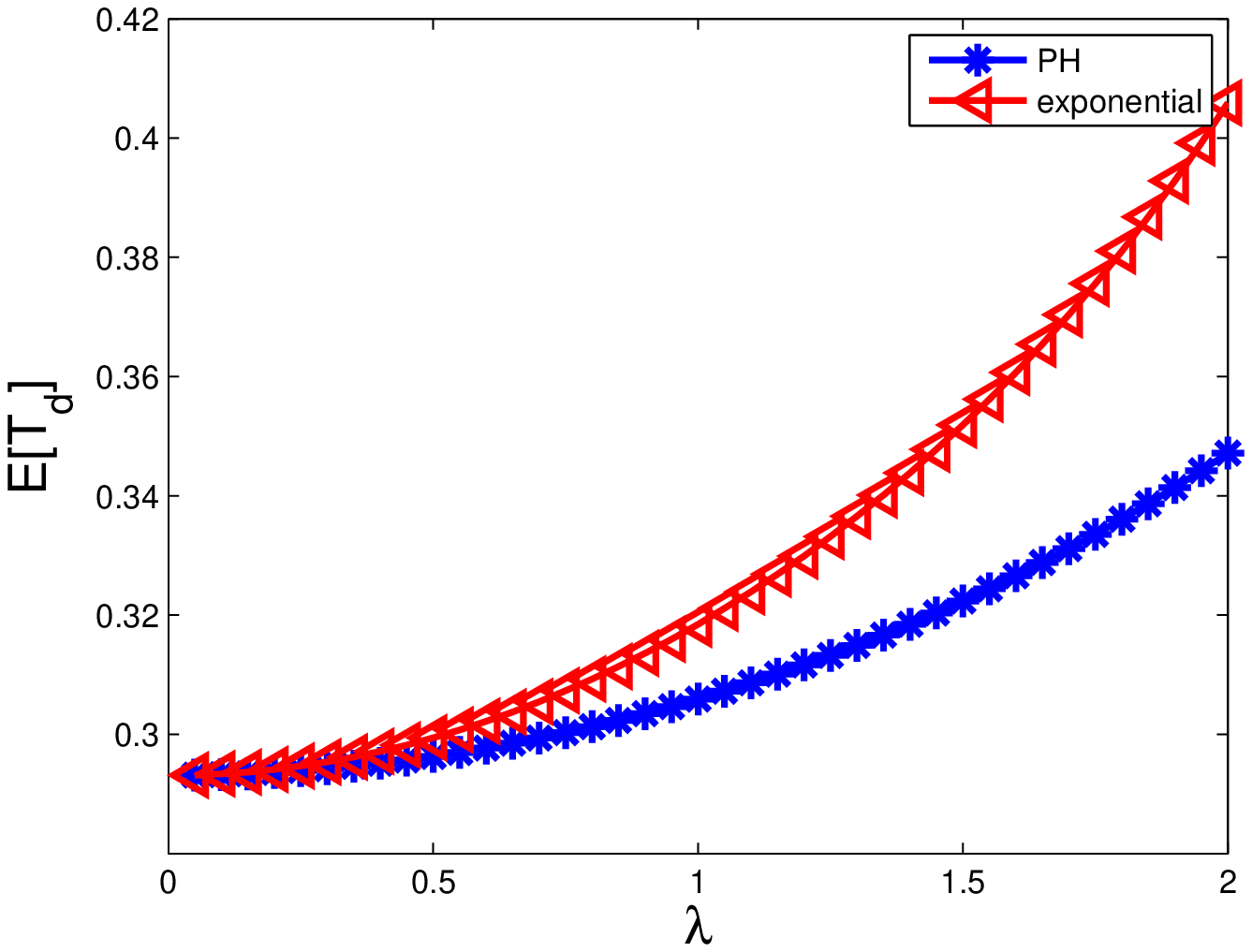}
\end{minipage}
\caption{$E\left[ T_{d}\right]$s of the PH and exponential distributions for
$T(1)$ and $T(2)$, respectively}
\end{figure}

For $m=3,d=5,\alpha \left( 1\right) =\left( 1/3,1/3,1/3\right) $ and $\alpha
\left( 2\right) =\left( 1/12,7/12,1/3\right) $,%
\begin{equation*}
T=\left(
\begin{array}{ccc}
-10 & 2 & 4 \\
3 & -7 & 4 \\
0 & 2 & -5%
\end{array}%
\right) .
\end{equation*}%
Table \ref{table: 5} shows how the doubly exponential solution ($\pi _{1}$
to $\pi _{4}$) depends on the vectors $\alpha \left( 1\right) $ and $\alpha
\left( 2\right) $, respectively.

\begin{table}[tbh]
\caption{The doubly exponential solution depends on the vectors $\protect%
\alpha$}
\label{table: 5}\centering {\scriptsize
\begin{tabular}{|c|c|c|}
\hline
& $\alpha=(\frac{1}{3},\frac{1}{3},\frac{1}{3})$ & $\alpha=(\frac{1}{12},%
\frac{7}{12},\frac{1}{3})$ \\ \hline\hline
$\pi_{1}$ & (0.0741,\;0.1358 ,\;0.2346) & (0.0602,\;0.1728,\;0.2531) \\
\hline
$\pi_{2}$ & (5.619e-05,\;1.030e-05,\; 1.779e-04 ) & (7.182e-05,\;2.063e-04,%
\;3.020e-04) \\ \hline
$\pi_{3}$ & (1.411e-20,\;2.587e-20,\;4.469e-20) & (1.739e-19,\;4.993e-19,%
\;7.311e-19) \\ \hline
$\pi_{4}$ & (1.410e-98,\;2.586e-98,\;4.466e-98) & (1.444e-92,\;4.148e-92,%
\;6.074e-92) \\ \hline
\end{tabular}
}
\end{table}

\section{Concluding remarks}

In this paper, we provide a matrix-analytic solution for supermarket models.
We describe the supermarket model with PH service times as a system of
differential vector equations, and provide a doubly exponential solution to
the fixed point of the system of differential vector equations. We also
provide some numerical examples to illustrate that our approach is effective
and efficient in the study of randomized load balancing schemes with
non-exponential service requirements, such as, Erlang service time
distributions, hyper-exponential service time distributions and PH service
time distributions. We expect that this approach will be applicable to study
other randomized load balancing schemes, for example, generalizing the
arrival process to non-Poisson such as renewal process or Markovian arrival
process, generalizing the service times to general probability
distributions, and analyzing retrial and processor-sharing service
disciplines.

%


\begin{thebibliography}{99}
\bibitem{Azar:1999} Y. Azar, A.Z. Broder, A.R. Karlin and E. Upfal (1999).
Balanced allocations. \textit{SIAM Journal on Computing} \textbf{29},
180--200. A preliminary version of this paper appeared in \textit{%
Proceedings of the Twenty-Sixth Annual ACM Symposium on the Theory of
Computing}, 1994.

\bibitem{Bra:2010} M. Bramson, Y. Lu and B. Prabhakar (2010). Randomized
load balancing with general service time distributions. In \textit{%
Proceedings of the ACM SIGMETRICS international conference on Measurement
and modeling of computer systems}, pages 275--286.

\bibitem{Dah:1999} M. Dahlin (1999). Interpreting stale load information.
\textit{IEEE Transactions on Parallel and Distributed Systems} \textbf{11},
1033 - 1047.

\bibitem{Eag:1986a} D.L. Eager, E.D. Lazokwska and J. Zahorjan (1986).
Adaptive load sharing in homogeneous distributed systems. \textit{IEEE
Transactions on Software Engineering} \textbf{12}, 662--675.

\bibitem{Eag:1986b} D.L. Eager, E.D. Lazokwska and J. Zahorjan (1986). A
comparison of receiver-initiated and sender-initiated adaptive load sharing.
\textit{Performance Evaluation Review} \textbf{6}, 53--68.

\bibitem{Eag:1988} D.L. Eager, E.D. Lazokwska and J. Zahorjan (1988). The
limited performance benefits of migrating active processes for load sharing.
\textit{Performance Evaluation Review} \textbf{16}, 63--72.

\bibitem{harchol97} M. Harchol-Balter, A.B. Downey (1997). Exploiting
process lifetime distributions for dynamic load balancing. \textit{ACM
Transactions on Computer Systems} \textbf{15}, 253--285.

\bibitem{Luc:2006} M. Luczak and C. McDiarmid (2006). On the maximum queue
length in the supermarket model. \textit{The Annals of Probability} \textbf{%
34}, 493--527.

\bibitem{Mar:2001} J.B. Martin (2001). Point processes in fast Jackson
networks. \textit{Annals of Applied Probability} \textbf{11}, 650-663.

\bibitem{Mar:1999} J.B. Martin and Y.M Suhov (1999). Fast Jackson networks.
\textit{Annals of Applied Probability} \textbf{9}, 854--870.

\bibitem{Mit:1996a} M.D. Mitzenmacher (1996). Load balancing and density
dependent jump Markov processes. In \textit{Proceedings of the
Thirty-Seventh Annual Symposium on Foundations of Computer Science}, pages
213--222.

\bibitem{Mit:1996b} M.D. Mitzenmacher (1996). The power of two choices in
randomized load balancing. PhD thesis, University of California at Berkeley,
Department of Computer Science, Berkeley, CA, 1996.

\bibitem{Mit:1998a} M. Mitzenmacher (1998). Analyses of load stealing models
using differential equations. In \textit{Proceedings of the Tenth ACM
Symposium on Parallel Algorithms and Architectures}, pages 212--221.

\bibitem{Mit:1999a} M. Mitzenmacher (1999). On the analysis of randomized
load balancing schemes. \textit{Theory of Computing Systems} \textbf{32},
361--386.

\bibitem{Mit:1999b} M. Mitzenmacher (1999). Studying balanced allocations
with differential equations. \textit{Combinatorics, Probability, and
Computing} \textbf{8}, 473--482.

\bibitem{Mit:2000} M. Mitzenmacher (2000). How useful is old information?
\textit{IEEE Transactions on Parallel and Distributed Systems} \textbf{11},
6--20.

\bibitem{Mit:2001} M. Mitzenmacher (2001). The power of two choices in
randomized load balancing. \textit{IEEE Transactions on Parallel and
Distributed Computing} \textbf{12}, 1094-1104.

\bibitem{Mit:2001a} M. Mitzenmacher, A. Richa, and R. Sitaraman (2001). The
power of two random choices: a survey of techniques and results. In \textit{%
Handbook of Randomized Computing: volume 1}, edited by P. Pardalos, S.
Rajasekaran and J. Rolim, pages 255-312.

\bibitem{Mit:1998} M. Mitzenmacher and B. V\"{o}cking (1998). The
asymptotics of selecting the shortest of two, improved. In \textit{%
Proceedings of the 37th Annual Allerton Conference on Communication,
Control, and Computing}, pages 326--327. A full version is available as
Harvard Computer Science TR-08-99.

\bibitem{Mir:1989} R. Mirchandaney, D. Towsley, and J.A. Stankovic (1989).
Analysis of the effects of delays on load sharing. \textit{IEEE Transactions
on Computers} \textbf{38}, 1513--1525.

\bibitem{Suh:2002} Y.M. Suhov and N.D. Vvedenskaya (2002). Fast Jackson
Networks with Dynamic Routing. \textit{Problems of Information Transmission}
\textbf{38}, 136\{153.

\bibitem{Tel:2002} M. Telek and A. Heindl (2002). Matching moments for
acyclic discrete and continuous phase-type distributions of second order.
\textit{International Journal of Simulation: Systems, Science \& Technology}
\textbf{3}, 47--57.

\bibitem{Voc:1999} B. V\"{o}cking (1999). How asymmetry helps load
balancing. In \textit{Proceedings of the Fortieth Annual Symposium on
Foundations of Computer Science}, pages 131--140.

\bibitem{Vve:1996} N.D. Vvedenskaya, R.L. Dobrushin and F.I. Karpelevich.
(1996). Queueing system with selection of the shortest of two queues: An
asymptotic approach. \textit{Problems of Information Transmissions} \textbf{%
32}, 20--34.

\bibitem{Vve:1997} N.D. Vvedenskaya and Y.M. Suhov (1997). Dobrushin's
mean-field approximation for a queue with dynamic routing. \textit{Markov
Processes and Related Fields} \textbf{3}, 493--526.

\bibitem{Zhou:1988} S. Zhou (1988). A trace-driven simulation study of
dynamic load balancing. \textit{IEEE Transactions on Software Engineering}
\textbf{14}, 1327--1341.
\end{thebibliography}
\end{document}